\documentclass[12pt,aps]{revtex4}

\begin{document}
\title
{"Slow Light" and "Slow Current"}
\author{Zapasskii V.S. and Kozlov G.G.}

\begin{abstract}
 It is shown that the effect of hole-burning under conditions of coherent
population oscillations as well as the light pulse delay in a
saturable absorber (a modification of the "slow light" effect) can
be interpreted, in a comprehensive way, in terms of {\it intensity
spectrum} of the light and {\it intensity-related susceptibility}
of the non-linear medium. The physical content of these effects is
illustrated by a simple electric circuit with a non-linear
resistor which realizes a full analog of the saturable absorber.
In this case the effect of hole-burning in the absorption
spectrum of the medium is converted in to the effect of
hole-burning in the  frequency dependence of resistance of the
non-linear resistor and the effect of "slow light" -- in to the
effect of "slow current".
\end{abstract}
\maketitle
\hskip20pt Institute of Physics, St.-Petersburg State
University,
 St.-Petersburg, 198504 Russia.
\hskip100pt {\it e}-mail:  gkozlov@photonics.phys.spbu.ru

\section{Introduction}

At the end of the last century, there have been performed remarkable experiments
\cite{Kasapi,Hau}, that
demonstrated the possibility to reduce  the group velocity of light in a medium
by many orders of magnitude. The idea of "slow light" has rapidly become
extremely popular,  gained a high prestige, and soon turned into a separate
direction of quantum optics.
In the first experiments on "slow light",
the anomalously narrow dispersion feature of the atomic medium
was created using the effect of electromagnetically induced transparency
\cite{Harris}.
Later on, a different scheme of strong reduction of the group velocity
of light has been proposed based on the so-called effect of coherent population
oscillations (CPO)
\cite{Bigelow1,Bigelow2}.  The classical CPO effect is observed with the use of
two monochromatic light beams, namely, the probe and the pump.
The pump beam creates the spectral hole detected by the probe  \cite{Schwarz}.
In the observations of the CPO-based "slow light", the pump beam is not explicitly
used, but, nevertheless, the time delay of the light pulse is ascribed  to the
 CPO-based hole-burning effect.
In the papers  \cite{Zapasskii,Selden1,Aleksandrov}, it has been shown
that all manifestations of the CPO-based "slow light"
perfectly match the simplest model of saturable absorber (see, e.g.,  \cite{Selden}),
which does not imply spectral hole-burning with appropriate modification of the  group
velocity of light.
Still, the studies on the CPO-based "slow light" proceed, with the new
interpretation of the pulse delay in a saturable absorber being regarded
as self-evident. In our opinion, the possibility of such a revision
of interpretation of a well-known effect of classical nonlinear optics
is related to insufficiently clear understanding of physical nature
of the hole-burning effect under conditions of the coherent population
oscillations.

The goal of this paper is to demonstrate, in the most straightforward way, the fact
that
all manifestations of "slow light" in a saturable absorber, as well as the CPO effect
are controlled exclusively by frequency dependence of its  {\it intensity-related}
response.
It is important to emphasize that in terms of the intensity spectra,
which allow one to describe this problem in a comprehensive way,
neither the susceptibility of the system, nor the optical perturbation
exhibit any spectral features at optical frequencies. Moreover, these effects
cannot be considered, by any means, as specifically optical.  As an example, we
present a simple electric circuit with a nonlinear resistor that realizes a full analogue
of the saturable absorber in the field of the light wave including the effect of hole-
burning in the absorption
spectrum (converted into the effect of hole-burning in the frequency dependence of
 of the resistor's resistance) and the effect of "slow light" (converted into the effect of "slow
current").
The proposed analogy makes it possible, in our opinion, to better understand physical
content of the effects under consideration.

In this paper, we will talk only about "slow" light, keeping in mind
that all the results are equally valid for the so-called "fast" light (the light
with a superluminal or even "negative" group velocity) upon sign inversion  of the
intensity-related nonlinearity
of the system (i.e., upon replacing the bleachable absorber by the inverse one).

\section{Saturable absorber: Basic properties}

By the saturable absorber is usually meant a layer of an optical medium
whose transmission  $\ae$ varies with intensity $I$ of the light, following, as a rule
with some delay, its variations.  As follows from the name of the saturable absorber, its absorption
exhibits saturation with
growing light intensity, though here this fact will be of no importance.
In the most general form, the  relation between intensities of the
incident  ($I$) and transmitted ($I_{out}$) light can be expressed by the formula

\begin{equation}
I_{out} = \ae(I,t) I
\end{equation}

and the dynamics of establishment of the equilibrium transmission
is given by the conventional relationship

\begin{equation}
 \dot{\ae}={\ae_{eq}-\ae\over\tau}
 \end{equation}

where $\tau$ is the relaxation time of the absorber.
To describe the effects under study, it suffices to restrict oneself
to the case of small variations of the light intensity $\delta I$
in the vicinity of $I_0$ and to consider
a "linear" saturable absorber, with its stationary transmissivity $\ae_{eq}$
linearly varying with $\delta I$:
\begin{equation}
\ae_{eq}(I)=\ae_0(I_0)-\ae_1 \delta I,
\end{equation}
In the framework of these simple assumptions, one can easily find the amplitude
and phase
characteristics of the intensity-related response of the saturable absorber (Fig. 1).
These curves represent frequency dependences of the amplitude  $|K(\omega\tau)|$
and phase
$\phi(\omega\tau)$ of the light intensity modulation
at the exit of the saturable absorber under condition of harmonic modulation of the
light intensity at the entrance.

As seen from the figure,
the spectrum of such an "intensity susceptibility"
displays a single feature lying in the range of low frequencies comparable with the
inverse relaxation time  $\tau$ (for more detail see  \cite{Zapasskii}).

We want to attract special attention to the fact that this approach uses only the
notion
of the light intensity (the field amplitude squared), whose spectrum has no spectral
peculiarities at optical frequencies. For this reason, this solution does not contain explicitly
either optical characteristics
of the light beam (spectral composition, degree of monochromaticity), or
spectral properties of the absorber at optical frequencies (structure of the
absorption spectrum,
the nature of the band broadening, etc.). At the same time, this model describes
comprehensively
the CPO effect, as well as all manifestations  of "slow light".

Indeed, the phase and amplitude characteristics of transmission for the intensity-
modulated light passing through a saturable absorber studied in the papers on the hole-burning
in a homogeneously broadened band and on "slow light" \cite{Hil,Mal,Bigelow1,Bigelow2,Baldit,Zhang},
precisely agree with the above model and do not require to invoke
the effect of hole-burning in the optical spectrum of the absorber.

The same can be said about the light pulse delay in the saturable absorber -- the
effect that is  regarded as the main evidence for the group velocity reduction and for the
"slow light" effect.
Since in the range of low modulation frequencies
 $\omega$, the phase delay  $\delta\phi$ linearly varies with frequency,
the time delay  $\delta t =\delta\phi/\omega)$
ceases to be frequency-dependent. As a result, the light pulse, whose intensity
spectrum
lies within the interval  $\delta\omega<<\tau^{-1}$,
displays practically pure temporal shift with no reshaping.

In the conventional CPO effect, the weak modulation of the light passing through
the saturable absorber is created using two monochromatic waves with
close frequencies, one of them strong (pump wave) and the other weak (probe
wave).
This intensity-modulated light in this case, evidently, exhibits all the
amplitude and phase changes mentioned above. But now, one detects variations in
the optical
spectrum (rather than in the intensity spectrum) of the transmitted light. This
possibility is provided by the fine spectral structure of the light.
Now, it becomes important that for the sufficiently low modulation frequencies
(for the probe light frequency being close enough to that of the pump), the
transmissivity
of the medium is modulated at the difference frequency, and the light passing
through such a modulator
acquires additional spectral components, with the frequency of one of them
being exactly coincident with that of the probe wave. As a result,
 the probe beam exhibits, in the vicinity of the pump wave
frequency,
an effective dip of the absorption spectrum (the appropriate spectral feature is also
displayed by
the phase behavior of the probe wave). It is this dip that is
considered to be responsible for all manifestations of "slow light" in a saturable
absorber.

In fact, we may conclude that physical content of the effects of
CPO and "slow light" is confined to transformation of the intensity
spectrum of the light transmitted through the saturable absorber in accordance  with
the transfer function shown in Fig. 1.

It should be noted that observation of the CPO effect, generally,
does not necessarily imply real population oscillations, because the mechanism of the
nonlinear
dependence (1) may be not related to the light-induced changes
in population of the initial state of the optical transition.
Such a dependence may result from, say, a temperature shift
of the optical spectrum that changes the overlap of the spectrum of the light beam
with the absorption spectrum of the medium with increasing pump
power, provided that the medium is heated due to nonradiative relaxation
of the excitation. In other words, the saturable absorber, under conditions of
the modulated pump, is capable of demonstrating coherent oscillations of
losses, rather than populations.  Moreover, this effect is, by no means,
specifically optical and can be observed on nonlinear systems of other physical
nature.

To get a clearer idea about the so-called effect of "hole-burning"
in the optical spectrum of a saturable absorber and about attendant  effects of
"slow light",
let us turn to a simple nonlinear system that has nothing to do with optics but
realizes an isomorphic model of the saturable absorber.

\section{"Slow current"}

Consider the electric circuit shown in Fig. 2. The key element of the
circuit (which actually simulates the saturable absorber) is the resistor $R$,
whose resistance varies with temperature which, in turn, follows
(with some temporal delay) the power released on the resistor.
By the input signal, we will mean the voltage $U$ applied to the circuit and
by the output, the voltage  $U_{out}$ falling on the resistor  $\rho$ (proportional
to the current in the circuit):
\begin{equation}
U_{out}={\rho\over R+\rho} U\equiv \ae U
\end{equation}

Here, the transfer coefficient $\ae$ is the counterpart
of transmissivity of the saturable absorber (see Eq. (1)).
The dynamics of this coefficient is controlled by the equation
\begin{equation}
 \dot{\ae}={\ae_{eq}-\ae\over\tau}
 \end{equation}
 Here, $\tau$ is the relaxation time and $\ae_{eq}$
is the equilibrium value of the transfer coefficient.

We will assume that for small variations of the input voltage  $\delta U$
in the vicinity of some  value  $U_0$, the equilibrium value of the coefficient
$\ae_{eq}$
varies linearly with  $\delta U$:
\begin{equation}
\ae_{eq}(U)=\ae_0(U_0)-\ae_1 \delta U,
\end{equation}
One can easily see that the relation between the input and output voltage,
in this case, is completely equivalent to that between intensities of the input and
output light in the saturable absorber (see Eqs. (1--3)), and the  quadripole
  described above realizes
analog of the saturable absorber in the field of the light wave. Indeed,
the amplitude and phase transformations of the input voltage modulation, in this
circuit, are controlled by the same transfer functions (Fig. 1), whereas a
sufficiently long pulse (in the scale of inverse relaxation times) of the input voltage
modulation will exhibit a time delay, thus demonstrating the effect of "slow current"
(if the resistance falls with increasing temperature, then the pulse of the current
is delayed with respect to the voltage). Note that
the fullness of the analogy is not hindered by the absence of the
high-frequency carrier, which excludes any possible role of the "hole-burning"
effect.

As one can easily see, however, no essential changes occur if
the dc modulated voltage is replaced by a ac modulated voltage by filling
the envelope with a high-frequency carrier. In this case, all specific
features of the low-frequency response of the quadripole  to the modulation of the
effective voltage will remain the same (including the effect of "slow current"),
but simultaneously there will appear the possibility to analyze transformation of
the spectra of the high-frequency carrier by "slow" nonlinearity of the system.
 Using this circuit, one can easily  make the experiment
similar to that described in classical papers on observation
of the hole-burning effect under conditions of the coherent population
oscillations. For this purpose, one should apply to the input of the circuit
two high-frequency signals with close frequencies (the "pump" voltage and the
"probe" voltage), and detect at the exit  only the weak "probe" voltage.
 In this kind of experiment, one will observe  a narrow peak of current
in the range of the "probe" frequencies close to that of the "pump".
This peak can be interpreted  as the effect of "hole-burning" in the frequency
dependence of resistance of the nonlinear resistor  $R$.

Figure 2 (a--d) schematically shows characteristic types of transformation
of a modulated signal by the above electric circuit.

\section{Conclusions}

In this paper, we have made an attempt to vividly demonstrate
the fact that the effect of coherent population oscillations, as well
as  all manifestations of "slow light" result exceptionally from
the "resonance" of the intensity-related susceptibility
of the absorber at zero frequency. This spectral feature of the intensity
susceptibility is really capable of modifying spectral composition
of the light passing through the absorber (to the extent of modification of its
intensity spectrum),
which is revealed in the effect of "hole-burning" in the homogeneously broadened
absorption
spectrum. However, the question about possibility to treat this hole as a usual
spectral
feature of a linear absorption spectrum and, in particular, to apply
to it the Kramers-Kronig relations needs special consideration.
Specifically, as we already pointed out in  \cite{Zapasskii},
the possibility to detect the "hole" using standard absorption spectroscopy  may
depend on the internal phase structure of the light beam.
In any case, we are not aware of experiments in which the effect
of the pulse delay was observed in the presence of the pump beam. As for the
"slow light" effect observed with a single light pulse, which, according to
\cite{Bigelow1}, combines the functions of the pump and probe beams, we see no grounds, in this case,
to invoke the effect of coherent population oscillations for explanation of the
pulse delay.
We believe that the physical nature of the effects considered above is convincingly
demonstrated  by the proposed analogy between the saturable absorber
 and the nonlinear resistor in the electric circuit. This circuit
simulates the effects of "slow" and "fast" light (depending on the sign of the
temperature nonlinearity of the resistor), as well as the above effect of "hole-burning". This
model simultaneously shows that the light pulse delay is not connected with effects of
propagation and that it is inappropriate to invoke the notion of
group velocity to interpret the effect.

\begin{figure}
\caption{The amplitude and phase  spectra of intensity susceptibility of the saturable
absorber. }
\end{figure}
\begin{figure}
\caption{ Electrical analog of the saturable absorber and the main types of
transformations of the modulated voltage without (a and b) and with (c and d)
the carrier frequency. The dashed lines on the right plots show time dependences of
the input voltage.}
\end{figure}
\end{document}